%% file: main.tex
\documentclass[sigconf, nonacm]{acmart}

\newcommand\vldbdoi{XX.XX/XXX.XX}
\newcommand\vldbpages{XXX-XXX}
\newcommand\vldbvolume{14}
\newcommand\vldbissue{1}
\newcommand\vldbyear{2020}
\newcommand\vldbauthors{\authors}
\newcommand\vldbtitle{\shorttitle} 
\newcommand\vldbavailabilityurl{https://github.com/celeris-labs/oasis}
\newcommand\vldbpagestyle{plain}

\usepackage{enumitem}

\usepackage[framemethod=default]{mdframed}

\usepackage{xcolor}
\definecolor{odcodebg}{HTML}{F6F5F2}

\newmdenv[
  backgroundcolor=odcodebg,
  linewidth=0pt,
  hidealllines=true,
  roundcorner=0pt,
  innerleftmargin=2pt,
  innerrightmargin=2pt,
  innertopmargin=1pt,
  innerbottommargin=1pt,
  skipabove=0pt,
  skipbelow=0pt,
  nobreak=true
]{codebox}

\usepackage{listings}
\definecolor{odpurple}{HTML}{A626A4} 
\definecolor{odblue}{HTML}{207AFA}
\definecolor{odorange}{HTML}{3C7A3B}

\makeatletter
\newcommand{\odnum}[1]{{\ifnum\lst@mode=\lst@Pmode\relax\color{odpurple}\fi#1}}
\makeatother

\lstdefinestyle{onedarksql}{
  language=SQL,
  basicstyle=\ttfamily\color{black},
  keywordstyle=\color{black},
  deletekeywords={SELECT,FROM,WHERE,AS,CAST,AND,OR,NOT,ON,USING,GROUP,BY,ORDER,HAVING,JOIN,INNER,OUTER,LEFT,RIGHT,BETWEEN,IN,IS,LIKE},
  morekeywords=[2]{SELECT,FROM,WHERE,AS,CAST,AND,OR,NOT,ON,USING,
                   GROUP,BY,ORDER,HAVING,LIMIT,JOIN,INNER,OUTER,LEFT,RIGHT,
                   EXPLAIN,ANALYZE,BETWEEN,IN,IS,LIKE},
  keywordstyle=[2]\bfseries\color{odblue},
  stringstyle=\color{odorange},
  literate={0}{{\odnum{0}}}1 {1}{{\odnum{1}}}1 {2}{{\odnum{2}}}1
           {3}{{\odnum{3}}}1 {4}{{\odnum{4}}}1 {5}{{\odnum{5}}}1
           {6}{{\odnum{6}}}1 {7}{{\odnum{7}}}1 {8}{{\odnum{8}}}1
           {9}{{\odnum{9}}}1,
  numbers=left,
  numberstyle=\ttfamily\scriptsize\color{black},
  numbersep=6pt,
  xleftmargin=1.8 em,
  frame=none,
  showstringspaces=false,
  columns=fullflexible,
  keepspaces=true,
  breaklines=true,
  captionpos=b,
  aboveskip=4pt, belowskip=4pt
}

\usepackage[capitalise,nameinlink]{cleveref}
\crefname{section}{Section}{Sections}
\Crefname{section}{Section}{Sections}
\crefname{figure}{Figure}{Figures}
\Crefname{figure}{Figure}{Figures}
\crefname{table}{Table}{Tables}
\Crefname{table}{Table}{Tables}
\crefname{lstlisting}{List.}{List.}
\Crefname{lstlisting}{Listing}{Listings}

\newcommand{\labeltitle}[1]{\vskip 0.03in \noindent\emph{#1}.}

\newcommand{\eg}{e.\,g.,\ }
\newcommand{\ie}{i.\,e.,\ }

\begin{document}
\title{Oasis: Hiding the Cost of Querying Parquet Files in the Datapath}

\author{Jonas Dann}
\affiliation{%
  \institution{ETH Z\"urich}
  \city{Z\"urich}
  \country{Switzerland}
}
\email{jonas.dann@inf.ethz.ch}

\author{Luca Tagliavini}
\authornote{Work was done while the author was a student at ETH Z\"urich.}
\affiliation{%
  \institution{Apple}
  \city{Z\"urich}
  \country{Switzerland}
}
\email{l_tagliavini@apple.com}

\author{Gustavo Alonso}
\affiliation{%
  \institution{ETH Z\"urich}
  \city{Z\"urich}
  \country{Switzerland}
}
\email{alonso@inf.ethz.ch}

\begin{abstract}
Cloud-native database systems disaggregate compute and storage resources to improve cost efficiency over traditional monolithic architectures through elasticity and resource pooling.
Studies of production data warehouse workloads show that scans (including round trips to storage) account for roughly half of total query runtime.
Data lakes and lakehouses amplify this bottleneck through per-query decoding of storage-optimized, compressed file formats such as Parquet. 
As storage and network bandwidth continue to outpace CPU cost-performance, the CPU cycles spent on decoding increasingly undermine the cloud's cost-efficiency promise.
This has led to a wave of specialization across the stack with custom hardware at cloud-vendor scale at the extreme end. 
We build on this trend and present Oasis, a data-processing SmartNIC that offloads Parquet decoding into the network datapath as a custom hardware accelerator. 
Oasis features a hardware decoder architecture, software abstraction layer, and end-to-end integration with DuckDB. 
Our evaluation shows that Oasis hides the cost of Parquet decoding behind the network datapath with minimal overhead, overlapping the scan with the remainder of the query execution.
In the best case, this almost doubles DuckDB query throughput.
\end{abstract}

\maketitle

\pagestyle{\vldbpagestyle}
\begingroup\small\noindent\raggedright\textbf{PVLDB Reference Format:}\\
\vldbauthors. \vldbtitle. PVLDB, \vldbvolume(\vldbissue): \vldbpages, \vldbyear.\\
\href{https://doi.org/\vldbdoi}{doi:\vldbdoi}
\endgroup
\begingroup
\renewcommand\thefootnote{}\footnote{\noindent
This work is licensed under the Creative Commons BY-NC-ND 4.0 International License. Visit \url{https://creativecommons.org/licenses/by-nc-nd/4.0/} to view a copy of this license. For any use beyond those covered by this license, obtain permission by emailing \href{mailto:info@vldb.org}{info@vldb.org}. Copyright is held by the owner/author(s). Publication rights licensed to the VLDB Endowment. \\
\raggedright Proceedings of the VLDB Endowment, Vol. \vldbvolume, No. \vldbissue\ %
ISSN 2150-8097. \\
\href{https://doi.org/\vldbdoi}{doi:\vldbdoi} \\
}\addtocounter{footnote}{-1}\endgroup

\ifdefempty{\vldbavailabilityurl}{}{
\vspace{.3cm}
\begingroup\small\noindent\raggedright\textbf{PVLDB Artifact Availability:}\\
The source code, data, and/or other artifacts have been made available at \url{\vldbavailabilityurl}.
\endgroup
}

\input{sections/01_introduction}
\input{sections/02_background}
\input{sections/03_oasis}
\input{sections/04_software}
\input{sections/05_evaluation}
\input{sections/06_conclusion}

\begin{acks}
We would like to thank AMD for the donation of the Heterogeneous Accelerated Compute Cluster (HACC) at ETHZ which was used to obtain the experimental results presented in this paper.
A special thanks also goes out to Geert Roks for the support of the cluster and Maximilian Heer and Benjamin Ramhorst for their continued efforts on the SCENIC and Coyote open-source projects. 
This work was funded in part through an unrestricted grant from AMD.
\end{acks}


\bibliographystyle{ACM-Reference-Format}
\bibliography{main}

\end{document}

%% file: sections/01_introduction.tex
\section{Introduction}
\begin{figure}[bt]
	\centering
	\includegraphics[width=\linewidth]{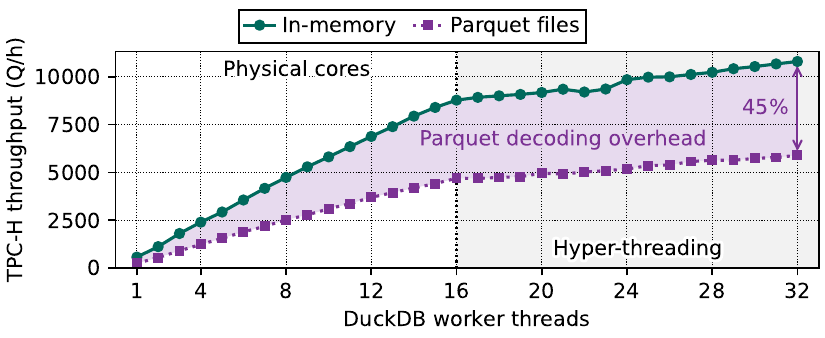}
	\caption{TPC-H (scale factor 30, four streams) throughput benchmark in queries per hour using DuckDB on pre-loaded data in memory vs on Parquet files across 1--32 DuckDB worker threads. Data decoding makes up almost half of the query runtime when querying Parquet files directly.}
	\label{fig:introduction}
\end{figure}
Cloud-native database systems have improved upon traditional, monolithic database architectures by disaggregating storage and compute, improving cost efficiency through elasticity and resource pooling.
However, disaggregated architectures introduce their own challenges around data movement from cloud object storage to compute nodes.
Recent studies of production data warehouse workloads show that scanning data alone accounts for roughly half of total query runtime~\cite{journals/pvldb/SzlangBCDFHOOM25, journals/pvldb/RenenHPVDNLSKK24}, rising to over 80\% for read-only queries~\cite{journals/pvldb/RenenL23}.
In data lakes and lakehouses \cite{conf/cidr/Zaharia0XA21}, which process queries directly on raw input files, this bottleneck is amplified by per-query decoding of encoded and compressed file formats such as Parquet~\cite{parquet}, which are optimized for storage rather than scan performance.
To quantify this effect, \cref{fig:introduction} shows that decoding Parquet files nearly halves DuckDB's~\cite{conf/sigmod/RaasveldtM19} query throughput compared to processing data in DuckDB's in-memory format.
This decoding tax is fundamentally a compute problem: every query pays it in CPU cycles, regardless of how fast the underlying storage delivers the data.

As storage and network bandwidths have increased by orders of magnitude over the last decade, the bottleneck for data processing in the cloud has shifted to the CPUs~\cite{journals/pacmmod/KuschewskiGNL24, journals/pvldb/ZengHSPMZ23, conf/nsdi/VuppalapatiMATM20, conf/cidr/HuB0025}.
Specifically, the stagnant cost-performance of CPUs undermines the cost-efficiency promise of the cloud.
Enabled by the scale of the cloud, these economics have triggered a wave of specialization across the whole stack, from compiled query engines~\cite{journals/pvldb/Neumann11, journals/pvldb/KerstenLKNPB18} through unikernels at the operating-system level~\cite{conf/eurosys/KuenzerBLSJGSLT21, journals/pvldb/LeisD24} to GPU query processing~\cite{conf/sigmod/InterlandiBHCSL26} and custom accelerators~\cite{conf/nsdi/FirestonePMCDAA18, conf/micro/KarandikarLKZPN21} at the hardware level. 
The common thread is specialization that reduces the overhead introduced by general abstractions (such as general-purpose CPUs).
In the case of Parquet decoding, dedicated accelerator prototypes~\cite{conf/icfpt/PeltenburgLH0AH20, conf/damon/KwonIRMF26} have demonstrated the potential for hardware specialization.
However, these accelerators lack a clear deployment path, do not cover the subset of the Parquet standard used in practice, and are isolated efforts without end-to-end integration into a database system.

SmartNICs, which co-locate user-accessible processing capabilities with the network interface controller (NIC), offer precisely such a deployment path: they can process data directly in the network datapath from object storage to compute~\cite{journals/access/KfouryCMAGC24}.
Cloud providers already deploy SmartNICs at massive scale: for over a decade, Microsoft Azure has equipped all new servers with FPGA-based SmartNICs to offload the overhead of tasks such as network virtualization~\cite{conf/nsdi/FirestonePMCDAA18}.
Similarly, AWS Nitro~\cite{aws2022nitro} and Alibaba CIPU~\cite{alibaba2022cipu} offload virtualization, storage access, and security functions across their respective fleets.
While these deployments target cloud infrastructure, the same hardware is now being enabled for database systems: Microsoft Fabric has recently begun targeting FPGAs for query execution~\cite{conf/sigmod/InterlandiBHCSL26}.
Inspired by this, we propose Oasis, a data-processing SmartNIC for cloud-native database systems that offloads line-rate Parquet decoding into the network datapath and integrates end-to-end as a DuckDB extension. 
For the custom hardware design backing Oasis, we doubled the decompression throughput of a popular Snappy decompression core and designed a hardware decoder pipeline that can reach line-rate decoding in many cases.
Our experimental results show that we can completely hide the cost of decoding Parquet files behind the network datapath of the database system with minimal overhead.
Our key contributions are:

\begin{enumerate}[label=(C\arabic*)]
    \item \textbf{Oasis SmartNIC} architecture (\cref{sec:oasis}) that sits on the network datapath of the compute node and hosts a set of hardware Parquet decoders (\cref{sec:decoder}).
    \item \textbf{Software abstraction layer} for Oasis, handling software-side output buffer management, decoder configuration, and scheduling of jobs to hardware decoders (\cref{sec:sal}).
    \item \textbf{DuckDB extension} with end-to-end integration of Oasis into the DuckDB database system (\cref{sec:extension}).
\end{enumerate}

%% file: sections/02_background.tex
\section{Background}
In this section, we introduce Parquet and its encodings, Snappy, and hardware design considerations for on-datapath SmartNICs.

\subsection{Parquet File Format}
\begin{figure}[bt]
	\centering
	\includegraphics[width=\linewidth]{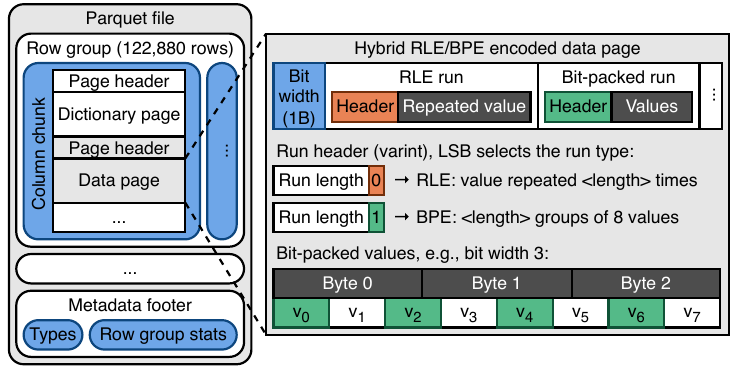}
	\caption{Parquet compressed, columnar file format horizontally partitioning data into row groups and details on hybrid run-length/bit-packing encoding for data pages.}
	\label{fig:parquet}
\end{figure}

Apache Parquet~\cite{parquet} is an open-source columnar file format designed for efficient, compressed storage of tabular data.
\cref{fig:parquet} shows the structure of a Parquet file.
At the top level, the file is divided horizontally into one or more row groups, each containing a contiguous subset of the table's rows (for example, rows $0$--$122{,}879$ under the default row group size of the DuckDB Parquet writer).
Within each row group, the data of each column is stored as a column chunk, which is in turn divided into pages.
Parquet compresses data in two steps: page data is first encoded, and then a general-purpose compression scheme, such as Snappy, is applied to each page of a column chunk independently.
There are several different encodings, but the three used in practice~\cite{conf/damon/KwonIRMF26} are plain encoding, dictionary encoding, and hybrid run-length (RLE)/bit-packing (BPE) encoding.
Thus, there are two page types: data pages and dictionary pages.

\labeltitle{Dictionary encoding}
A dictionary page written at the start of the column chunk records the distinct values of the column, and subsequent data pages store indices into that dictionary rather than the values themselves. 
This is very effective for low-cardinality columns since the resulting indices into the dictionary are dense and thus occupy only a small value range that can be represented by a much smaller number of bits than the original data. 
In practice, DuckDB encodes columns with a dictionary until the dictionary grows too large, exceeding around $1\,\mathrm{MiB}$, and then falls back to plain encoding for the remainder of the data.

\labeltitle{Hybrid RLE/BPE} The indices produced by dictionary encoding are then encoded as hybrid RLE/BPE pages.
The scheme interleaves RLE and BPE runs, choosing whichever representation is more compact for a given run of values.
Each run is prefixed by a variable-length integer (varint) header whose least-significant bit selects the run type.
If the bit is clear, the run is run-length encoded: the remaining header bits give the repetition count, and the payload stores the repeated value, using $\lceil \text{bit width}/8 \rceil$ bytes.
If the bit is set, the run is bit-packed: the remaining header bits give the run length in groups of eight values, and the payload stores the values back-to-back at a fixed bit width, least-significant bits first.
Notably, the bit width itself is not encoded in the run headers but is derived from the dictionary size and written as a single byte preceding the stream of indices.
When writing Parquet files, DuckDB commits to an RLE run once four or more consecutive values are equal.
Otherwise, it accumulates values into bit-packed blocks of $256$ values.

\labeltitle{Metadata} Parquet stores metadata at three levels.
At the page level, each data page is preceded by a page header recording the page type, the number of values, byte length, and encoding information.
Each page also optionally has Dremel-style~\cite{journals/cacm/MelnikGLRSTV11} definition and repetition levels that track nullability and nested structures, respectively.
At the column chunk level, the column metadata records the column's type, the byte offsets of the first data page and the optional dictionary page, value and null counts, and optional min/max statistics.
The compression scheme is also defined per column chunk.
At the file level, a footer at the end of the file contains the complete file metadata: the schema, the number of rows, and the row-group and column-chunk metadata for every column chunk.
The footer length is stored in the last bytes of the file and must be read first.

\begin{figure*}[t]
	\centering
	\includegraphics[width=.9\linewidth]{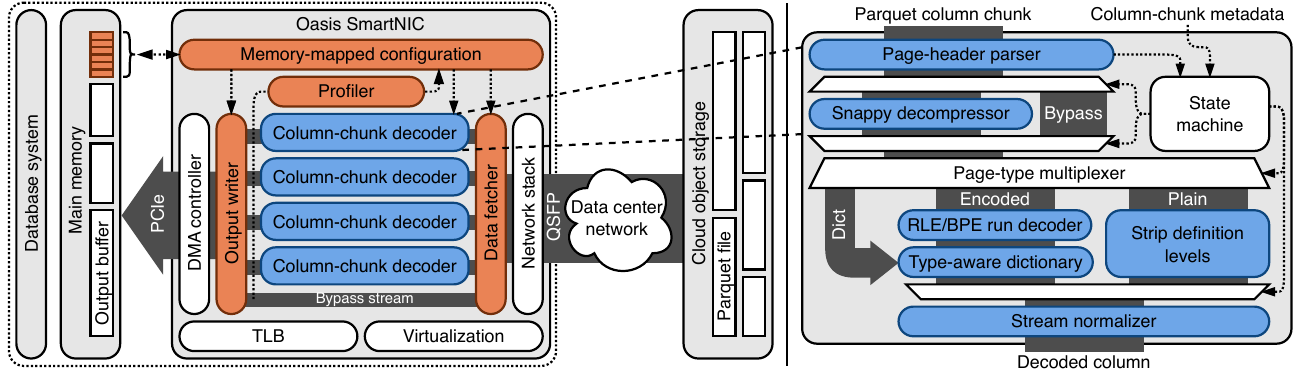}
	\caption{Oasis SmartNIC architecture and detailed hardware column-chunk decoder pipeline.}
	\label{fig:column-chunk-decoder}
\end{figure*}

\subsection{Snappy Compression Scheme}
Snappy~\cite{snappy} is a dictionary-based compression scheme that prioritizes decompression speed over compression ratio.
A Snappy-compressed page begins with a preamble encoding the uncompressed length as a varint, followed by a sequence of elements of two kinds: literals, which copy raw bytes from the compressed stream to the output, and copies, which duplicate a run of bytes that appeared at an offset in the already-decompressed output.
Each element starts with a tag byte whose two least-significant bits identify the element type: $00$ denotes a literal, while $01$, $10$, and $11$ denote copies with one-, two-, and four-byte offsets, respectively.
For literals, the remaining six bits of the tag byte encode lengths of up to $60$ bytes directly.
Longer lengths are stored in up to four additional bytes following the tag.
Copies with a one-byte offset encode lengths of $4$--$11$ bytes and offsets of up to $2{,}047$ bytes, packing three offset bits into the tag byte, whereas copies with two- and four-byte offsets encode lengths of up to $64$ bytes with the offset stored in the subsequent bytes.
The compressor processes its input in blocks of $64\,\mathrm{KiB}$, so copy offsets never exceed this distance and the working set of the decompressor remains small.

\subsection{SmartNIC Hardware Design Considerations}
\label{sec:fpga}
SmartNICs are widely deployed in the cloud~\cite{conf/nsdi/FirestonePMCDAA18} and have become an established target for offloading data-intensive
processing~\cite{conf/sigmod/Faghih0IB24, journals/pvldb/GiouroukisNPZM25}.
We prototype our Oasis SmartNIC on a network-enabled field-programmable gate array (FPGA).
FPGAs map custom digital circuit designs (a set of logic gates and their connections) to a grid of resources (\eg look-up tables, registers, and SRAM) connected with a programmable interconnection network.
Synthesizing a design for such an architecture involves assigning the logic gates and other structures to their physical equivalents on the chip (\emph{place}) and allocating interconnect wires for their connections (\emph{route}).
Unlike in instruction-based processors (CPUs and GPUs), the performance of a hardware design is not limited by the number of instructions that can be executed in each clock cycle but by (i) how efficiently the available hardware resources can be utilized to exploit data and pipeline parallelism and (ii) the achievable clock frequency of the design.
Specifically, the achievable clock frequency is restricted by the longest path (measured in nanoseconds) between two registers, which we call the critical path.
The critical path needs to fit inside one clock cycle (\eg $1/250\,\mathrm{MHz} = 4\,\mathrm{ns}$).
The path length is given by the sum of signal propagation delays on the path and the time that the gates on the path need to stabilize their output after a signal arrives at their input.
In complex hardware designs, the number of logic gates in the paths and the fan-out of signals limit the clock frequency.
Less often, the general density of the design becomes the limiting factor for routing because it restricts freedom in placing the logic.
To optimize clock frequency, we try to either restructure the hardware design from scratch or pull the logic apart with more pipelining, which, however, is not always possible (\eg if there are tight feedback loops for backpressure).
The performance of the design is linear in the clock frequency that the synthesis achieves.
All of this leads to very different design principles for data processing in specialized hardware that we will explain in more detail when we apply them to our Oasis SmartNIC.

%% file: sections/03_oasis.tex
\section{Oasis SmartNIC: Overview}
\label{sec:oasis}
Oasis is an on-datapath SmartNIC that sits on the network datapath of the compute node in a disaggregated architecture with direct connections to the network through its own network port on one side and to the main memory of the host system through PCIe on the other side. 
Data that the database system fetches from cloud object storage always flows through the Oasis SmartNIC.
\cref{fig:column-chunk-decoder} shows the Oasis architecture, with $n$ column-chunk decoders at its heart, each decoding one column chunk of a Parquet file at a time.
The figure also shows a detailed datapath of the hardware column-chunk decoder pipelines that we discuss in \cref{sec:decoder}.
Integration with the database system is discussed in \cref{sec:integration}.

The Oasis SmartNIC is controlled by the database system through a memory-mapped configuration that can be read and written from the host CPU.
The high-level datapath is made up of a network stack handling the network protocol-level details of fetching data from a remote storage server, a data fetcher that instructs the network stack to fetch data based on configurations written through the memory-mapped interface, a set of column-chunk decoder pipelines, an output writer that coalesces data from the column-chunk decoders and generates memory requests to write the data to main memory, and a direct-memory access (DMA) controller that handles the details of writing data through PCIe to main memory.
There is also a bypass stream that we use to fetch metadata and string columns, which we currently do not handle in hardware and instead decode on the CPU.
Beyond the datapath, profiling logic sitting on the output streams counts clock cycles of the data flow, a translation lookaside buffer (TLB) performs address translations from virtual to physical addresses, and further virtualization components support multiple parallel streams, interrupts, and related functionality.

The key property of this architecture is that data is decoded as it streams into the database system: a column chunk is decompressed and decoded while it streams from the network port to host memory, rather than being written to memory in its encoded form first and then read back by the CPU and decoded.
Communication and decoding overlap at cache-line ($64\,\mathrm{bytes}$) granularity.
Oasis further exploits the parallelism the Parquet format inherently offers: column chunks are self-contained units carrying their own metadata and dictionary, so the data fetcher can hand independent column chunks to the $n$ instantiated column-chunk decoders without any synchronization between them.
Three benefits follow.
First, we eliminate redundant copies of the encoded data.
Second, the decoding overhead is lifted off the CPU entirely and can be fully hidden by overlapping query execution with decoding.
Third, decoding throughput stops being bound by how many instructions we can execute on the CPU and is instead bound only by how efficiently we can build the decoding pipeline and by the network and interconnect speeds.
The data that arrives in the output buffers is already decoded, in the representation that the database system expects, so no conversion step is left for the host and the handoff is zero-copy.

\section{Hardware Column-chunk Decoder}
\label{sec:decoder}
An Oasis hardware column-chunk decoder (right side of \cref{fig:column-chunk-decoder}) is a pipeline that extracts metadata from page headers, decompresses Snappy-compressed pages, and then decodes them.
Dictionary pages are forwarded to the value port of a type-aware dictionary, hybrid RLE/BPE-encoded pages go through a run decoder and the lookup port of the dictionary, and plain pages pass through a bypass that strips the definition levels.
The fully decoded values are then joined into one normalized stream that is written to a contiguous buffer. 
In the following, we describe the major components of the hardware column-chunk decoder pipeline.

\subsection{Page-header Parser}
The first module in the pipeline is the page-header parser, which decodes Apache Thrift \cite{slee2007thrift} encoded metadata that precedes every page in a column chunk.
The page-header parser extracts the number of expected values that are encoded in the page, the page type (dictionary or data), the encoding used on the page (plain or hybrid RLE/BPE encoding + dictionary), and whether it is the last page in the column chunk.
The last-page flag is used to flush the whole column-chunk decoder to reset it for the next column chunk.
The page-header parser is implemented in two stages as a state machine based on the Thrift schema of Parquet page headers.
In the first (slow) parsing stage, we decode the page header one encoded value at a time.
The varint encoding used on most values in the Thrift schema makes it very hard to decode values in parallel.
However, since the page header is a fraction of the data of a page, doing the header parsing iteratively instead of in parallel does not significantly impact performance.
In the second (fast) payload stage, we then pipe through the remaining bytes of the page, which are the actual payload.
The parsing stage uses a shift register that is shifted by one byte at a time (to reduce the multiplexer complexity) and one varint decoder placed at the first five bytes of that shift register.
We also maintain a skip register.
In each state of the state machine, if the skip register is 0, we decode a value.
We set the value of the skip register to the length of the decoded varint and transition to the next state.
We repeat this process until we reach the payload.
In the payload stage, we first flush the remaining bytes from the shift register and then forward full data beats to the output.

\subsection{Snappy Decompressor}
\begin{figure}[bt]
	\centering
	\includegraphics[width=\linewidth]{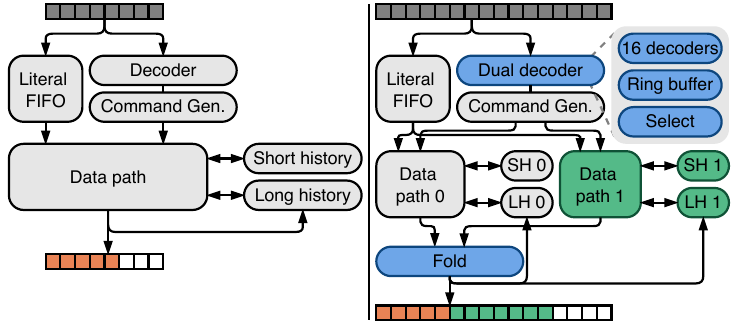}
	\caption{Original 8\,byte-wide Snappy decompressor and our optimized 16\,byte-wide design with dual-issue copies.}
	\label{fig:snappy}
\end{figure}
Snappy decompression is the computationally heaviest operation in Parquet decoding and poses the most significant bottleneck in the column-chunk decoder.
We base our Snappy decompressor on the existing decompression core VHSNUnzip\footnote{\url{https://github.com/abs-tudelft/vhsnunzip}}, which, to the best of our knowledge, is the fastest open-source core available.
We apply two optimizations to the existing design to double its performance (cf. \cref{fig:snappy}).
We first double the datapath width from 8\,bytes to 16\,bytes, allowing the design to both consume more input and emit more output in each clock cycle.
As we will show in the microbenchmarks (\cref{sec:micro:snappy}), this alone doubles performance for literal-heavy (\ie mostly uncompressed) input data that is piped through the decompression module directly.
For copy-heavy (\ie highly compressed) input data, however, the original design is severely limited by only decompressing a single copy command per clock cycle, which the wider datapath does not help with. 
Thus, we additionally extend the original design to decompress two copies per clock cycle.
To handle decoding multiple copies per data beat, we duplicate the literal storage implemented as on-chip SRAM with two separate datapaths: one taken by the first copy command and one by the potential second copy command.
This allows us to double copy throughput at the cost of doubling the resource utilization of that part of the datapath.
The individual results of the two datapaths are folded into one unified output afterwards.
Introducing the data parallelism of copy commands created a new issue with timing of the hardware design, however.
It forces the design to find two copy commands in the input per clock cycle, which creates a long chain of varint decoders, doubling the length of the critical path of the design.
We thus designed a two-entry ring buffer which holds pre-decoded varints that are decoded in a separate pipeline step before we actually select the two run headers for copy commands.
This decouples the decoding and header extraction from the input data beat and relaxes timing of the critical path.

\subsection{Hybrid RLE/BPE Run Decoder}
The hybrid RLE/BPE run decoder strips off the definition levels, reads the bit width from the payload, and proceeds decoding the remaining stream of runs.
The run decoder has a target line rate of 16 decoded values per clock cycle in the average case which means we have to retire one RLE run or 16 bit-packed values per clock cycle.
Run payloads are not aligned to data beat boundaries, so the data needed to decode the current run may straddle two data beats.
We derive a worst-case bound: the value of an RLE run needs at most $\lceil \text{bit width} / 8 \rceil$ bytes (at most 4, since the specification caps the bit width at 32), and one BPE decoding step needs bit width${}\cdot N$ bytes, where $N$ is the number of output values per data beat divided by eight.
Thus, storing the current data beat (64 bytes) plus one additional data beat is provably sufficient for runs overlapping data beats, so the module keeps exactly two data beats in a register and tracks a data offset and a varint offset into it.
Whenever consuming a run advances the data offset past the end of the first data beat, the second beat is shifted down into the first half of the register, both offsets are reduced by 64, and the freed upper half receives the next incoming data beat.
Offsets thus grow monotonically within a fixed 128-byte window, and every run extraction is a simple part-select at a byte offset.
One important subtlety is that the shift takes effect at the next clock edge, so values captured in the same cycle, such as the varint decoder's input bytes, must be indexed with pre-shift offsets.
Both RLE and BPE sub-decoders have several cycles of latency between input and output because they are internally pipelined to keep the critical path short.
We therefore decouple the input-driving logic from the output: whenever a sub-decoder is fed, a selector value is pushed onto a FIFO.
The head of that FIFO configures the multiplexer connecting the sub-decoders to the output.
This way the state machine can advance to the next header while the previous run's values are still in flight, and outputs are guaranteed to appear in the correct sequence.

Wiring the varint decoder for the run header combinatorially from the registered data through the offset would create a very long path through a wide multiplexer.
Instead, the next varint offset is precomputed and the varint decoder's input registers are only populated once the data beat that actually contains the next header has been ingested.
This breaks the header decoding out of the critical path while still allowing the transition from one run to the next in a single clock cycle.
We constrain the number of output identifiers per data beat to a multiple of eight, which guarantees byte-aligned offsets after every decoding step.
Similarly, we exploit two properties of real-world Parquet writers: they cap the number of values per page at roughly 1\,Mi, which shrinks all value-count signals from 32 down to 21 bits, and they cap dictionaries at roughly 1\,MiB (we support 2\,MiB, i.e., $2^{19}$ 32\,bit entries), which shrinks all identifier and bit-width signals down to 19 bits.
In hardware, we are not constrained to data types that are a multiple of 8 bit wide so this saves a lot of resources.
Neither constraint sacrifices generality in practice: the design still decodes any Parquet file produced by real-world writers.

\subsection{Type-aware Dictionary}
\begin{figure}[bt]
	\centering
	\includegraphics[width=\linewidth]{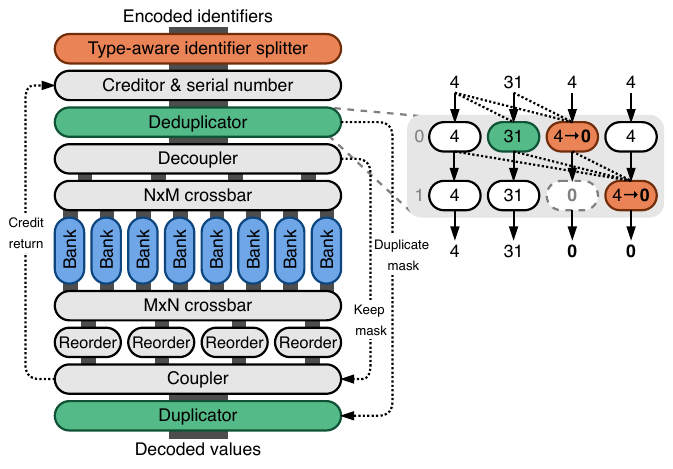}
	\caption{Type-aware dictionary with 8 banks and a detailed deduplicator pipeline for four ports.}
	\label{fig:typed-dict}
\end{figure}
The dictionary is the final stage of the hybrid-encoded page pipeline and is responsible for mapping the stream of identifiers produced by the run decoder back to the actual column values.
As the base for this step, we use the dictionary from the libSTF, a library of standard data processing components for specialized hardware\footnote{\url{https://github.com/fpgasystems/libstf}}, which is based on the crossbar and reordering design of Dann et al.~\cite{conf/fpl/Dann0F22}.
However, the original dictionary operates on fixed-width entries and has no notion of values whose width (32- or 64-bit) depends on the column's data type.
Since the dictionary is by far the heaviest component of the column-chunk decoder in terms of resources, we do not want to simply instantiate two of them to cover both value widths.
We propose a new type-aware dictionary with a fixed 64-byte datapath that extends the existing design with a type-aware identifier splitter (cf. \cref{fig:typed-dict}).
For 32-bit values, the mapping onto the underlying banks is trivial: each value occupies one cell in one bank, and each identifier retrieves exactly that cell.
For 64-bit values, we split each value into two 32-bit halves and store them in separate banks.
On lookup, each identifier $m$ is expanded into the two identifiers $2 \cdot m$ and $2 \cdot m + 1$, and the two cells retrieved from separate banks are concatenated to reconstruct the original 64-bit value.
Filling a full 64-byte output data beat with 32-bit values thus requires 16 identifiers per cycle, which is why the run decoder is specialized to produce 16 identifiers per data beat: with this parameterization, the dictionary can sustain full throughput for both data type widths.

Since dictionary-encoded columns typically have low cardinality, they frequently contain repeated identifiers after hybrid encoding, which are all routed to the same bank and serialize there, as each bank can only serve one request per cycle.
To avoid these bank conflicts, we add a deduplicator in front of the crossbar which compares every identifier lane of a data beat against all lanes preceding it.
If a lane matches an earlier lane, it is marked as a duplicate that is not forwarded to the crossbar.
The resulting duplicate mask, which stores the index of the first occurrence of the identifier, is forwarded to a duplicator at the end of the dictionary pipeline, where the duplicates are materialized.
The $\binom{16}{2} = 120$ pairwise comparisons required per data beat (we only show four lanes in \cref{fig:typed-dict}) are distributed over four pipeline stages such that each stage performs a roughly equal share of the comparisons, keeping the wide comparator logic off the critical path.
Deduplication thereby turns the worst-case conflict pattern---long runs of a single identifier, as commonly emitted by the run decoder---into a single bank lookup followed by a broadcast, so the dictionary sustains full throughput even on heavily skewed identifier streams.

\subsection{Limitations}
Oasis is able to decode column chunks written by the DuckDB and Arrow Parquet writers of any fixed-size 32-bit and 64-bit data types.
We do not yet support string columns or other compression schemes than Snappy.
However, decoding string columns will reuse large parts of the existing infrastructure including the decompressor and run decoder.
Only the dictionary would have to change for string decoding.
Similarly, other decompression cores (\eg for gzip or zstd) can be plugged into our architecture with only a light wrapper.
This has been shown to work in hardware in the past \cite{journals/pvldb/ChiosaMMAM22}.
Parquet defines a number of different encodings of which we support plain, dictionary, and hybrid RLE/BPE encoding.
A recent study of real-world data sets \cite{conf/damon/KwonIRMF26} shows that Parquet writers effectively exclusively use these encodings.
Additionally, we focus on non-nested relational tables without null values.

%% file: sections/04_software.tex
\section{End-to-end System Integration}
\label{sec:integration}
\begin{figure}[bt]
	\centering
	\includegraphics[width=\linewidth]{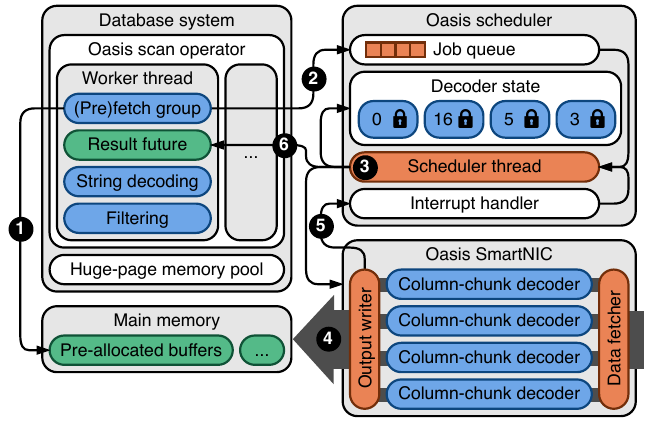}
	\caption{Control flow: worker threads submit decoding jobs to the Oasis scheduler, which schedules them onto the
	set of hardware decoders. Decoded data is written into pre-allocated host buffers for zero-copy handoff and passed back to the worker threads via interrupt-triggered result futures.}
	\label{fig:idea}
\end{figure}
This section describes how Oasis is integrated into a full database system. 
The central challenge is a mismatch in concurrency: a database executes queries with a large, potentially unbounded number of worker threads, while Oasis offers a small, fixed number of column-chunk decoders. 
The software stack bridges this gap while keeping the datapath zero-copy and the worker threads unblocked.

\subsection{Oasis Software Abstraction Layer}
\label{sec:sal}
The Oasis software abstraction layer (SAL) hides the accelerator behind an asynchronous submission interface.
It makes only two assumptions: scans are executed by parallel worker threads, and the database can suspend a task that waits on a future (systems that do not support this can block the worker thread instead).
Scanning a Parquet file proceeds in six steps (cf.\ Figure~\ref{fig:idea}). 
(1)~A worker thread claims a unit of work (\eg a Parquet row group) and pre-allocates an output buffer per projected column chunk from a pinned memory pool that is pre-mapped into Oasis's TLB.
The exact decoded size is known in advance based on the file metadata.
(2)~The worker thread then constructs a decoding job: a set of operator flows, one per column chunk, each a set registers of the memory-mapped configuration that must be written.
The registers include the file offset and size of the encoded column chunk that the data fetcher uses to fetch the input data, metadata about the column chunk such as the compression scheme and number of expected values, and the address of the pre-allocated buffer.
The worker then submits this decoding job to the Oasis scheduler job queue.
Submission is non-blocking: the job is placed on an unbounded queue, and the worker receives a result future.
To increase overlap of decoding and query processing, workers may prefetch multiple row groups ahead of consumption.
(3)~The scheduler thread dispatches queued flows onto hardware streams (explained in more detail in \cref{sec:scheduler}).
(4)~The hardware fetches, decodes, and writes results directly into the pre-allocated host buffers.
(5)~The hardware then raises one interrupt per completed buffer, encoding the stream id, the number of bytes written, and a last flag. 
The interrupt handler passes this information on to the scheduler.
(6)~The worker consumes buffers from its result future and maps them zero-copy into the database system's in-memory format.
A worker whose future is empty is descheduled through the database system's asynchronous task mechanism and woken by the interrupt-triggered completion.

\subsubsection{Memory-mapped Configuration}
All control flow uses configuration registers exposed over Oasis's memory-mapped PCIe control interface. 
The configuration register file is self-describing: a global configuration region advertises a system identifier and a table of 64-bit sub-configuration identifiers with their register address space bounds.
Each configuration register is backed by a FIFO of 64 slots.
At startup, the SAL enumerates this table and instantiates a configuration driver for each sub-configuration, with per-stream register groups inside each sub-configuration's address space (\eg the data fetcher exposes address, length, and RDMA queue pair registers per stream). 
One software binary thus drives any synthesized configuration regardless of decoder count, local or remote data fetchers, or optional profilers, and fails gracefully when a feature is absent from the hardware design.

\subsubsection{Oasis Scheduler}
\label{sec:scheduler}
The scheduler multiplexes the unbounded flow queues onto a fixed number of hardware streams with a greedy policy.
Flows are classified by the capability they require: \emph{decode} flows must run on a stream with a decoder, whereas \emph{bypass} flows must run on the bypass stream.
Each capability has its own queue, so a capability whose streams are saturated never head-of-line blocks the other.
Internally, the scheduler tracks per-stream state consisting of an occupancy counter, which records the number of flows currently enqueued on that stream, and a dispatch lock that serializes enqueuing flows.
On submission, the scheduler may take a fast path: if at least one stream of the corresponding capability is not saturated, the job's flows are dispatched directly to the least-loaded stream.
Otherwise, they are appended to the corresponding queue and dispatched later.
Whenever an interrupt arrives, the scheduler thread is woken up, decrements the occupancy counter of the corresponding stream and dispatches the oldest queued flow of the matching capability.
Since completions on one stream arrive in order, the interrupt handler matches each interrupt against the stream's FIFO of pending buffers and publishes the buffer on the result future of the owning job.
The last flow closes the future.
Greedy dispatch is sufficient because flows are independent and uniform in shape: fairness follows from the FIFO order of the queues, and load balance from always choosing the least-loaded stream.

\subsubsection{Stream Profiling}
Attributing bottlenecks inside the accelerator is difficult from the host, so each decoder carries a hardware stream profiler on its input and output. 
Per stream, the profiler counts handshake cycles, starved cycles (the consumer waits for data), stalled cycles (the producer is back-pressured), and idle cycles between consecutive transfers on a stream. 
Counting starts with the first handshake and trailing idle time is excluded, so the counters cover exactly the active window. 
The counters are read destructively through configuration registers, meaning they are reset once read. 
This lets us distinguish, per decoder, whether utilization is lost to data starvation or to back-pressure from the output.

\subsection{Oasis DuckDB Extension}
\label{sec:extension}
Finally, we describe the DuckDB extension for our Oasis SmartNIC that integrates it end-to-end into the query processing of DuckDB as a table function (\ie scan operator).
We also describe the RDMA server that we use to expose the Parquet files over the network. 
In future work, we will add support for S3-compatible storage APIs through HTTP.
For metadata operations and the DuckDB Parquet extension baseline, we also implement an RDMA file system based on the DuckDB file system extension hook.

\subsubsection{\texttt{read\_oasis()} Table Function}
\begin{figure}[t]
\small
\begin{codebox}
\begin{lstlisting}[style=onedarksql]
SELECT sum(l_extendedprice * l_discount) AS revenue
FROM read_oasis('rdma://lineitem.parquet')
WHERE l_shipdate >= CAST('1994-01-01' AS date)
  AND l_shipdate <  CAST('1995-01-01' AS date)
  AND l_discount BETWEEN 0.05 AND 0.07
  AND l_quantity < 24;
\end{lstlisting}
\end{codebox}
\caption{TPC-H benchmark query 6 using the Oasis extension's \texttt{read\_oasis()} table function.}
\label{lst:tpch-q6}
\end{figure}
The Oasis extension exposes SmartNIC-accelerated scanning of Parquet files as a new table function called \texttt{read\_oasis()}, a drop-in replacement for DuckDB's \texttt{read\_parquet()} that can be used in arbitrary SQL queries (cf. \Cref{lst:tpch-q6}) in place of regular in-memory tables.
Table functions in DuckDB return metadata such as the table schema and cardinality estimates during bind time (after the query is parsed, before it is planned).
Thus, at bind time, our table function reads the Parquet footer through the bypass stream of Oasis and reports the bind information to the planner. 
Our table function supports projection pushdown, filter pushdown, and filter pruning, so only referenced columns are fetched and decoded, filter-only columns are dropped from the operator's output, and dynamic filters produced by hash joins at run time are merged into the pushed-down filter set before the first row group is scanned. 
Pushed-down predicates are applied at three levels: row groups whose column statistics evaluate a predicate as always-false are skipped entirely, predicates that the statistics prove always-true are excluded from the runtime filter set, and the remainder is evaluated on the decoded vectors.
The table function plugs into DuckDB's morsel-driven parallelism: each worker thread claims row groups from a shared cursor and keeps a configurable number of them in flight (prefetching), submitting each as an independent decoding job to the Oasis scheduler with one decode flow per fixed-width column and one bypass flow per string column. 
While the FPGA decodes, the worker is never blocked: if the head group's results are not yet ready, the scan returns an asynchronous task and DuckDB reschedules it once the hardware signals completion through the result future. 
Decoded column buffers are handed to the engine zero-copy and are sliced in lockstep into $2{,}048$-row vectors.
Output vectors point directly to the Oasis-written buffers, whose lifetimes are tied to the vectors by shared ownership.
Queries that consume no column values (e.g., \texttt{COUNT(*)}) are answered directly from the Parquet metadata.

Variable-length columns are not yet supported by the hardware column-chunk decoders.
We thus fall back to DuckDB's native Parquet column readers on the CPU for string columns. 
To avoid synchronous remote reads on this path, the compressed bytes of string column chunks are fetched through the same decoding job as bypass flows and staged in host memory, where the CPU readers consume them. 
String predicates are evaluated during decoding, for dictionary-encoded pages once per dictionary entry rather
than once per row.
Following the late-materialization order of the filter pipeline, the remaining string columns are decoded only for rows that survive all predicates.
Pages whose rows are entirely filtered out are skipped without ever being read.

\subsubsection{RDMA Server}
The RDMA server exposes a set of Parquet files as a single memory region through RDMA to as many queue pairs as the client side establishes.
The raw bytes of the Parquet files are packed back-to-back and prefixed with a directory of file names, offsets, and lengths, so a client can bootstrap by reading the directory and then issue reads for arbitrary byte ranges of any file by translating them into offsets within the region.  
Connection setup follows an out-of-band queue-pair metadata exchange over TCP.
The client opens one connection per read-request stream, and the server creates a new queue pair per connection, serving several queue pairs concurrently against the shared region. 
After setup, the server's CPU is not involved in the datapath: all transfers are one-sided RDMA reads issued by the client and served entirely by the NIC, while the TCP connection is retained only as a liveness signal to tear down the queue pair when the client disconnects.

\subsubsection{RDMA Virtual File System}
On the client side, the Oasis DuckDB extension registers a read-only virtual file system with DuckDB that resolves paths under the \texttt{rdma://} scheme against the RDMA server's memory region.
The existing Parquet machinery (globbing, footer and metadata parsing, column-chunk reads) works on remote files unchanged. 
On first access, the file system establishes the RDMA queue pairs (one for each stream) and bootstraps the file directory once: it reads the first eight bytes of the region to learn the directory size, fetches the directory in a single transfer, and parses it into an in-memory map from file name to region offset and length. 
Opening a file then merely amounts to a local lookup in this map, and a file handle translates file-relative byte ranges into offsets within the remote region. 
Notably, our Oasis SmartNIC is always the RDMA endpoint on the client side, not a separate host NIC: every read is issued by Oasis as a one-sided RDMA read, and reads for the software (e.g., Parquet footers and metadata) are routed still through the bypass stream. 
As in other virtual file systems in DuckDB, we also coalesce column-chunk reads based on a tunable maximum byte gap between reads.

%% file: sections/05_evaluation.tex
\section{Experimental Evaluation}
\label{sec:evaluation}

We implement the Oasis SmartNIC as an FPGA prototype based on the SCENIC stream computation-enhanced SmartNIC \cite{journals/corr/abs-2604-15128}, Coyote v2 FPGA operating system \cite{conf/sosp/RamhorstKHDLA25}, and RoCE BALBOA data center RDMA stack \cite{journals/corr/abs-2507-20412}.
Oasis is embedded into Coyote as a virtual FPGA (vFPGA) and utilizes the network stack, virtual memory management, memory-mapped configuration registers, and interrupt interface.
The source and benchmark code is open source\footnote{\url{https://github.com/celeris-labs/oasis}} and can be used with any FPGA supported by Coyote v2.

\subsection{Experimental Setup}

\labeltitle{System Setup} 
For our experiments, as the compute node, we work with a server equipped with an AMD Alveo U55C FPGA board with a 100G network port plugged into the host server via PCIe v3.
The system features an AMD EPYC 7302P 16-core CPU with 32 hyper threads clocked at a maximum of 3.0GHz and 64GB of memory.
We use a server equipped with the same hardware plus a Mellanox ConnectX-6 100G RDMA NIC as the storage server.
Both servers are connected through a 100G switch.
To synthesize the hardware, we use Vivado 2024.2 with the \texttt{BUILD\_OPT} timing optimization flag enabled in Coyote.
We use DuckDB pinned to commit hash \texttt{15d3f23} which added Parquet async I/O and do not have to modify DuckDB in any way.
Everything for the Oasis integration is implemented as a DuckDB extension that can be loaded into any DuckDB instance of that version.
All software is compiled with GCC 11.4.0 and \texttt{CMAKE\_BUILD\_TYPE} set to \texttt{Release}.

\begin{table}[t]
\caption{Resource utilization of the Oasis SmartNIC (4 column-chunk decoders) on an AMD Alveo U55C FPGA.}
\label{tab:resource_utilization}
\centering
\footnotesize
\setlength{\tabcolsep}{3pt}
\begin{tabular}{l r r r r}
 Component & LUTs & Registers & BRAM & URAM \\
 \hline
 \hline
 Col.-chunk decoder (4x) & 84.4K (6.47\%) & 123K (4.73\%) & 16 (0.79\%) & 72 (7.50\%) \\
 \quad Page-header parser & 5718 (0.44\%) & 4199 (0.16\%) & 0 (0.00\%) & 0 (0.00\%) \\
 \quad Snappy decompressor & 10.6K (0.82\%) & 9060 (0.35\%) & 0 (0.00\%) & 8 (0.83\%) \\
 \quad Run decoder & 17.0K (1.31\%) & 3458 (0.13\%) & 6 (0.30\%) & 0 (0.00\%) \\
 \quad Type-aware dictionary & 42.8K (3.29\%) & 94.5K (3.63\%) & 10 (0.50\%) & 64 (6.67\%) \\
 \quad Stream normalizer & 3525 (0.27\%) & 4714 (0.18\%) & 0 (0.00\%) & 0 (0.00\%) \\
 \quad Other & 4622 (0.35\%) & 7310 (0.28\%) & 0 (0.00\%) & 0 (0.00\%) \\
 Other (Config, I/O, ...) & 26.2K (2.01\%) & 46.7K (1.79\%) & 85 (4.22\%) & 45 (4.69\%) \\
 Coyote shell & 338K (25.9\%) & 586K (22.5\%) & 479 (23.7\%) & 54 (5.62\%) \\
 \quad RDMA stack & 198K (15.2\%) & 382K (14.7\%) & 250 (12.4\%) & 54 (5.62\%) \\
 \hline
 Overall & 701K (53.8\%) & 1125K (43.2\%) & 628 (31.1\%) & 387 (40.3\%) \\
\end{tabular}
\end{table}

\labeltitle{Resource Utilization} Hardware designs grow spatially in the resources that they physically occupy on the chip while adding more and more functionality.
\cref{tab:resource_utilization} shows the resource utilization of the Oasis SmartNIC for a design with four hardware column-chunk decoders.
The three largest components of the column-chunk decoder are the decompressor, run decoder, and type-aware dictionary.
The type-aware dictionary specifically is the largest component because of its crossbar components that distribute the encoded identifiers to the dictionary banks and the >1\,MiB on-chip memory per instance implemented mostly as URAMs that store the dictionary.
The other resources used by the Oasis design are mostly from the output writer and configuration components.
The Coyote shell, specifically the RDMA datapath, takes up another major part of the hardware design.
Overall, the design takes up less than 50\% of the resources of the chip leaving ample room for more functionality.

\labeltitle{Data Sets} To generate the data sets, we use the TPC-H extension of DuckDB and write the resulting tables to Parquet files with the default DuckDB Parquet writer settings.
DuckDB emits row groups of $122{,}880$ rows, splitting purely by row count with no byte-based limit.
By default, all pages are Snappy compressed.
For each column chunk, the writer first builds a dictionary over all values of the row group and uses dictionary encoding unless the number of distinct values exceeds one fifth of the row group's rows, in which case it falls back to plain encoding.
Dictionary indices are written with $\lceil\log_2 d\rceil$ bits for a $d$-entry dictionary using the hybrid RLE/BPE scheme: a value repeated at least $4$ times is emitted as an RLE run (with unbounded run length), while shorter runs are accumulated into bit-packed blocks of $256$ values.
Data pages are only split once they reach $100\,\mathrm{MiB}$ of uncompressed data, so at this row group size each column chunk in practice contains a single data page.
We also tried TPC-H files written with the Arrow Parquet writer, but those achieve worse overall query throughput in DuckDB because Arrow's default row groups of $2^{20}$ rows are roughly $8.5\times$ larger than DuckDB's, yielding far less parallelism for decoding and query execution.
In general, Oasis supports files written by both writers.

\subsection{Microbenchmarks}

We first evaluate the performance of the hardware design and its optimizations through microbenchmarks with synthetic data.

\subsubsection{Snappy Decompressor Optimizations}
\label{sec:micro:snappy}
\begin{figure}[bt]
	\centering
	\includegraphics[width=\linewidth]{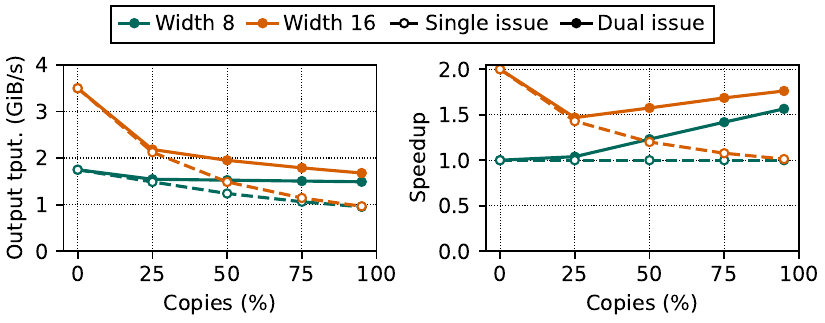}
	\caption{Snappy decompressor throughput and speedup for 8- and 16-byte datapath and single and dual issue copies.}
	\label{fig:micro:snappy}
\end{figure}
\cref{fig:micro:snappy} shows the different optimization stages of the Snappy decompression module on synthetic data with a varying fraction of literals to copies.
The baseline retires a single token per clock cycle over an 8-byte datapath.
We independently widen the datapath to 16\,bytes and add dual issue of copies, and report speedup over the baseline at the same copy fraction.
The two optimizations address opposite ends of the copy fraction scale. 
On pure literal input, throughput is limited by how many bytes the datapath moves per cycle, so doubling the width to 16\,bytes doubles throughput from $1.7$ to $3.5$\,GiB/s, while dual issue is inconsequential and the two curves coincide. 
At almost 100\,\% copies, the limit is instead the token rate: an average copy is far shorter than the datapath is wide, so the wide datapath idles and the single-issue 16-byte variant falls back onto the
baseline at $1.0$\,GiB/s. 
Here, retiring two copies per clock cycle increases throughput by $1.6\times$ at a width of 8\,bytes and $1.8\times$ at 16\,bytes.
Consequently, each optimization on its own leaves a region of the input distribution in which it contributes nothing and the corresponding curves decay to the baseline. 
Combined, they are complementary and yield at least $1.45\times$ over the whole range. 
The minimum at a copy fraction of $0.25$ is where neither bottleneck dominates: the literal path is no longer
saturated, but copies are still too sparse for dual issue to pay off. 
The fully optimized module also degrades most gracefully in absolute terms, sustaining at least $1.6$\,GiB/s across all input distributions, whereas the baseline drops to $1.0$\,GiB/s.

\subsubsection{Hybrid Run Decoding}
\begin{figure}[bt]
	\centering
	\includegraphics[width=\linewidth]{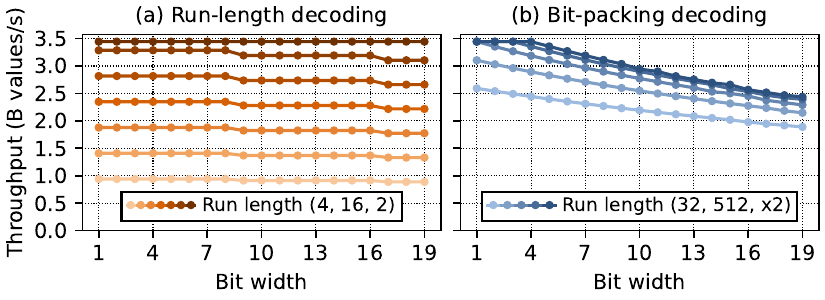}
	\caption{Hybrid run decoder throughput in billions of values per second for run-length decoding and bit-packing decoding for varying bit widths and run lengths.}
	\label{fig:micro:run-decoder}
\end{figure}
\cref{fig:micro:run-decoder} shows the throughput of the hybrid run decoder in billions of values per second measured on the output of the hybrid run decoder for run-length decoding and bit-packing decoding separately. 
We sweep bit widths from 1 to 19 bits, which cover the whole value range practically possible for 1\,MiB dictionaries.
We vary the run length between 4 and 16 values in steps of two for run-length encoded runs and between 32 and 512 values in powers of two for bit-packed runs.
DuckDB does not create smaller RLE runs than 4 values and packs 256 values into each BPE run.
For run-length decoding in \cref{fig:micro:run-decoder}(a), throughput is largely independent of the bit width across the entire sweep. 
This is expected since a run-length encoded run carries exactly one value that is replicated by the decoder, so the bit width only changes how many bits are consumed on the input side and never how many values are produced per clock cycle. 
Instead, throughput scales linearly with the run length from around 0.9 billion values per second at a run length of 4 to around 3.4 billion values per second at a run length of 16, since the decoder retires one run per clock cycle and the run length directly determines how many output values that run expands to. 
The scaling flattens for the longest runs where we reach the width of the output interface and further increasing the run length can physically not produce more values per clock cycle.
For bit-packing decoding in \cref{fig:micro:run-decoder}(b), the picture is inverted. 
Here, every value has to be extracted individually from the packed input, so the bit width directly determines how many values we can unpack per clock cycle and throughput decreases monotonically from around 3.2 billion values per second at a bit width of 4 to around 2.4 billion values per second at a bit width of 19. 
The run length has a much weaker effect than for run-length decoding and mainly amortizes the per-run overhead of consuming the run header and filling the pipeline. 
Consequently, the curves converge for run lengths of 128 values and above, while only the shortest bit-packed runs of 32 and 64 values pay a significant penalty of roughly half a billion values per second.
Increasing the BPE run length in the DuckDB Parquet writer would not yield higher throughput.
Our design is well tuned for it.
Overall, the hybrid run decoder sustains between 2 and 3.5 billion values per second across the relevant parameter space, and the bottleneck shifts from the run rate for short run-length encoded runs to the output interface for long ones and to the unpacking of individual values for wide bit-packed runs.

\subsubsection{Type-aware Dictionary}
\begin{figure}[bt]
	\centering
	\includegraphics[width=\linewidth]{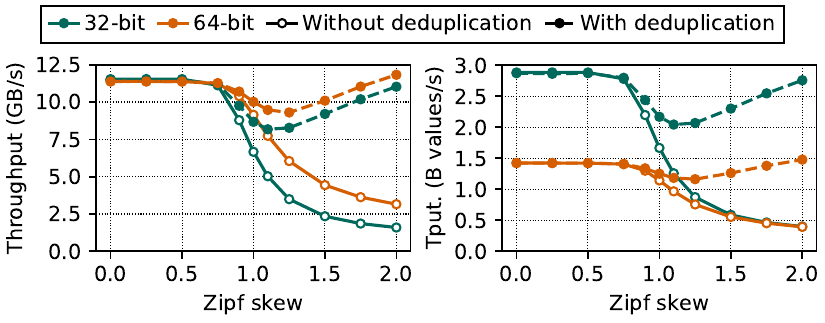}
	\caption{Type-aware dictionary throughput in gigabytes per second and billions of values per second for 32- and 64-bit values with and without deduplication under varying skew.}
	\label{fig:micro:dict}
\end{figure}
\cref{fig:micro:dict} shows the throughput of the type-aware dictionary in gigabytes per second and in billions of values per second for 32- and 64-bit values with and without deduplication of dictionary identifiers. 
We again measure on the output of the dictionary and count only the non-idle clock cycles. 
We generate the dictionary indices from a Zipf distribution and sweep the skew parameter from 0.0, i.e., a uniform access pattern, to 2.0, where a small number of dictionary entries accounts for almost all accesses, to cover the range from unique-heavy to highly repetitive columns.
Up to a skew of around 0.75, the dictionary runs at a constant 11.4 gigabytes per second independent of the value type, which is exactly the property we designed it for with the capability to tolerate localized bank conflicts. 
The datapath is limited by its width and not by the number of values it resolves, so the 32-bit configuration sustains 2.9 billion values per second where the 64-bit configuration sustains 1.4 billion values per second at the same byte rate.
Beyond a skew of around 0.75, accesses start to concentrate on a few hot dictionary entries and, without deduplication, the resulting bank conflicts have to be serialized. 
Throughput then collapses to 1.6 gigabytes per second for 32-bit and 3.1 gigabytes per second for 64-bit values at a skew of 2.0. 
The 32-bit configuration degrades more severely in bytes per second because it resolves twice as many values per clock cycle and therefore produces twice as many conflicting accesses.
This is also why both configurations converge to around 0.4 billion values per second in the right plot of \cref{fig:micro:dict}: once we are conflict bound, the limit is the rate at which we can serialize accesses per value and is no longer a function of the value width.
With deduplication, the same conflicts are resolved by collapsing repeated indices into a single dictionary access instead of serializing them. 
The worst case moves to a skew of around 1.25 with 8.2 gigabytes per second for 32-bit and 9.3 gigabytes per second for 64-bit values, since at intermediate skew a batch still contains several distinct hot values that cannot be merged into one access. 
For higher skew, throughput recovers because almost all values in a batch are identical and collapse into a single access. 
The bottleneck of the dictionary therefore shifts between the datapath width for low-skew data and the serialization of conflicting accesses for high-skew data, and deduplication bounds the worst case to a moderate dip instead of a collapse.

\subsubsection{Per-column Decoding Throughput}
\begin{figure}[bt]
	\centering
	\includegraphics[width=\linewidth]{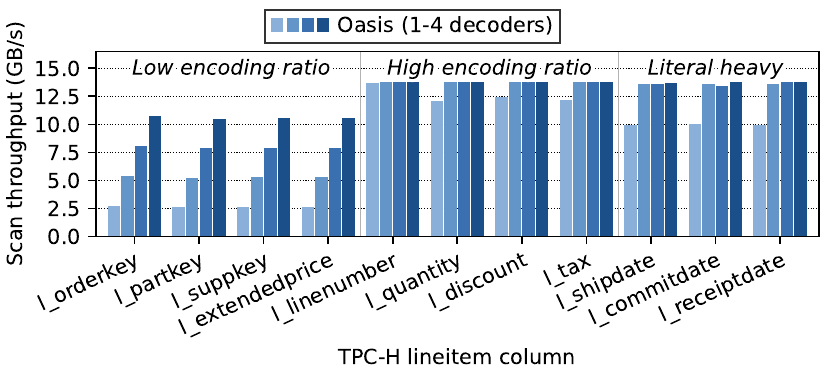}
	\caption{TPC-H (scale factor 30) lineitem table per-column decoding throughput in gigabytes per second of our Oasis SmartNIC using 1-4 hardware column-chunk decoders.}
	\label{fig:micro:columns}
\end{figure}
In our last microbenchmark, \cref{fig:micro:columns} shows per-column decoding throughput measured at the output of the hardware decoders.
The measurement counts only the clock cycles in which we actually transport data, excluding the idle cycles between transfers, to isolate raw decoding performance.
We use the non-string columns of the lineitem table as representation columns across the TPC-H benchmark.
In general, we are limited by the PCIe interconnect available on this hardware.
We observe three classes of columns characterized by their input data properties and the resulting performance.
First, the key columns and the price column of lineitem are Snappy-compressed with a high fraction of copies but have a low encoding ratio.
A single hardware decoder is therefore limited to the throughput of the Snappy decompressor, since the data does not grow in the subsequent decoding stages of the datapath.
However, performance scales perfectly across hardware decoders.
Second, the line number, quantity, discount, and tax columns also contain many copies in the Snappy-compressed input, but have a high encoding ratio because they hold relatively few unique values. 
The Snappy decompressor remains the slowest stage, but decoding inflates the data enough that a single hardware decoder almost saturates the PCIe link.
Third, literal-heavy columns such as the date columns are essentially uncompressed, so the Snappy decompressor passes the data through at full line rate and is no longer the bottleneck. 
The bottleneck of the hardware decoders thus shifts between the Snappy decompressor and the PCIe interconnect depending on the characteristics of the input data, which is why we scale to multiple decoder instances.

\subsection{End-to-end Scan Performance}
We now show the end-to-end performance of the Oasis extension scan operator on the TPC-H benchmark in a per-query runtime and throughput benchmark setting.

\subsubsection{Per-query Scan Runtime}
\begin{figure*}[t]
	\centering
	\includegraphics[width=\linewidth]{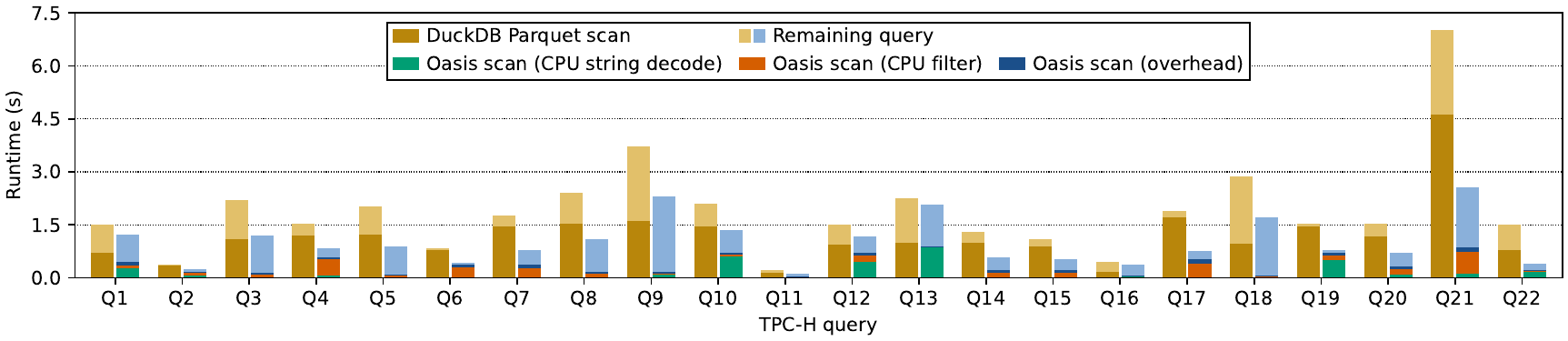}
	\caption{TPC-H (scale factor 30) benchmark per-query runtime in seconds comparing the baseline DuckDB Parquet extension to our Oasis SmartNIC using 4 hardware decoders both with 8 DuckDB worker threads. The remaining query runtime, excluding the scans, stays the same while the scanning can be overlapped with the query runtime.}
	\label{fig:queries}
\end{figure*}
In our first end-to-end experiment, we look at scan operator performance for each of the 22 TPC-H queries compared to the Parquet scan operator of the DuckDB Parquet extension.
DuckDB is configured to use 8 worker threads and four hardware decoders when using Oasis.
Again, we use Snappy compressed Parquet files written by DuckDB with default settings with scale factor 30 of the TPC-H data set.
The data is stored on the remote storage described above and is accessed through RDMA.
We warm the Parquet metadata cache of DuckDB beforehand to avoid network roundtrips for metadata fetching from skewing the end-to-end runtime, disable external file caching, and enable profiling so we can extract per-operator runtimes.
Each measurement point is the mean of three measurements.

In \cref{fig:queries}, we show the results with the TPC-H queries on the x-axis and the query runtime in seconds on the y-axis.
We are most interested in the relative runtimes of the baseline on the left of each query group compared to Oasis runtime on the right.
For the baseline, we split up the query runtime into the scan and remaining query runtime.
The scan runtime includes pushed-down filters.
As expected, we see that the remaining query runtime remains the same for all queries.
For the scan runtime itself, we see stark improvements cutting the total query runtime in half for a lot of queries.
For queries such as Q3, Q5, Q8, Q9, Q11, Q16, and Q18, we fully overlap the scan runtime with the remaining query runtime to fully hide the scan in the network datapath.
Only a small overhead remains for setting up the output buffers and the control flow.

For the other queries, we observe two major remaining overheads in the scan runtime.
For queries Q1, Q10, Q13, Q19, and Q22, we still see significant runtime for string decoding.
Two queries stick out in particular.
Q10 is pronounced because we decode five large string columns: c\_name, c\_phone, c\_address, c\_comment, and l\_returnflag.
For Q13, we decode the largest string column o\_comment and apply a \texttt{NOT LIKE} filter which counts towards the string decoding time.
The second major overhead in the scan operators is pushed-down filters for queries Q4, Q6, Q7, Q17, Q20, and Q21.
For Q4, the filter runtime is dominated by a pushed-down dynamic range filter on l\_orderkey and optional Bloom filter from the subsequent join.
We see purely static filters for Q6 where the query is a single filtered aggregation over lineitem, and evaluating the three pushed-down predicates on l\_shipdate, l\_discount, and l\_quantity makes up $79\%$ of its scan runtime.
Q17 is the opposite: its lineitem scan has only dynamic range and Bloom filters on l\_partkey created by the highly selective scan of part, where the predicates on p\_brand and p\_container retain only ${\sim}0.1\%$ of the rows.
Finally, Q21 accumulates the largest absolute filter runtime of all queries because its \texttt{EXISTS} and \texttt{NOT EXISTS} subqueries scan lineitem three times, each time evaluating dynamic filters on l\_orderkey and l\_suppkey produced by the joins.
Overall, we can hide almost the whole scan operator runtime in the network datapath except the pushed down filters and string decoding.
These now take up $84\%$ of the remaining scan runtime or summed across the total query runtime, ${\sim}15\%$ each across all queries.

\subsubsection{DuckDB Query Throughput}
\begin{figure}[bt]
	\centering
	\includegraphics[width=\linewidth]{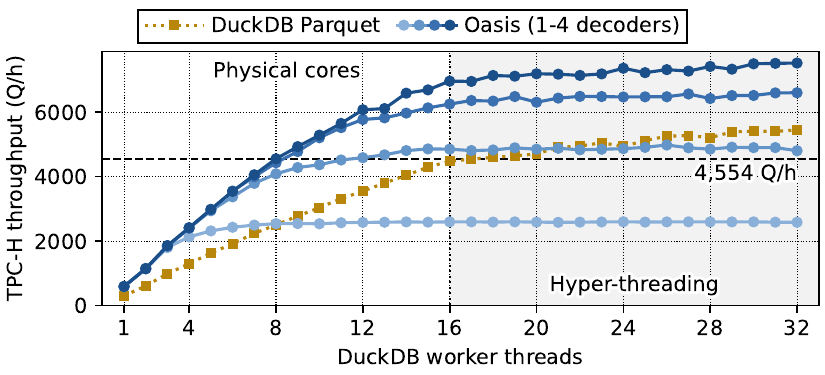}
	\caption{TPC-H (scale factor 30, four streams) throughput benchmark in queries per hour of the baseline DuckDB Parquet extension compared to our Oasis SmartNIC.}
	\label{fig:tpch}
\end{figure}
In our second end-to-end experiment we measure query throughput as per the TPC-H throughput benchmark.
We again use the same TPC-H scale factor 30 Parquet files on our remote storage server accessed through RDMA.
We also again warm the Parquet metadata cache of DuckDB beforehand and disable external file caching.
We restrict the number of worker threads with DuckDB to $n$ and also pin the execution to the first $n$ cores/hyper-threads of the CPU with \texttt{numactl} because we noticed that just setting the number of threads in DuckDB is not reliable.
The benchmark sets up four connections (\ie streams) to DuckDB each then submitting a randomly permuted sequence of the 22 TPC-H queries in parallel.
\cref{fig:tpch} shows the TPC-H throughput in queries per hour on the y-axis while scaling the number of worker threads on the x-axis of DuckDB's Parquet extension compared to our Oasis extension with 1--4 hardware column-chunk decoders. 

The Parquet extension baseline scales very similarly to the measurement run on local Parquet files from \cref{fig:introduction} because of the low overhead of RDMA and async I/O support in the DuckDB Parquet extension that is able to hide the network latency well.
Oasis scales very well, almost matching in-memory performance, up to 8 threads where it shows almost double the throughput compared to the DuckDB baseline at $4{,}554$ queries per hour.
After that, performance starts to level off because we are hitting the limit of the PCIe 3 interconnect with the decoded data.
We do not see this for the DuckDB baseline because the data is only decoded after crossing this bottleneck.
There, we never saturate the network bandwidth or PCIe interconnect which have roughly similar line rates.
However, Oasis with four hardware column-chunk decoders is still 40\% faster at the top end.
Regarding scaling, we are also restricted by PCIe interconnect.
One to two hardware decoders scales almost perfectly and we would expect the same for three and four decoders.
Oasis therefore returns half of the cores to actual query processing rather than using them for decoding and with fast interconnects, potentially fully gets rid of the decoding overhead matching in-memory performance on Parquet files through the network.

\subsubsection{Prefetching \& Decoder Utilization}
\begin{figure}[bt]
	\centering
	\includegraphics[width=\linewidth]{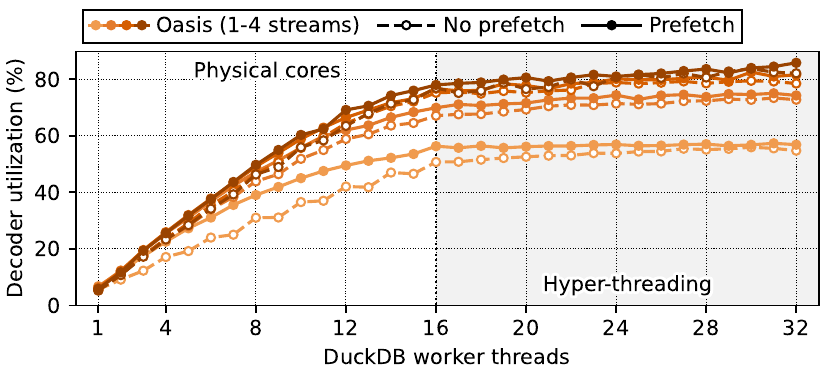}
	\caption{TPC-H (scale factor 30) throughput benchmark column-chunk decoder utilization for 1-4 streams of queries for Oasis SmartNIC with four hardware decoders.}
	\label{fig:utilization}
\end{figure}
Finally, we take a look at how prefetching influences column-chunk decoder utilization in Oasis.
We again run TPC-H throughput benchmarks at scale factor 30 with one to four streams submitting queries.
\cref{fig:utilization} shows the decoder utilization on the y-axis scaling across DuckDB worker threads on the x-axis for each number of streams without prefetching (\ie async I/O but prefetching depth set to 1) and prefetching enabled (prefetching depth set to 2).
Decoder utilization is calculated from the clock cycle counters of the stream profilers that sit on the output port of the decoders as the number of non-idle clock cycles relative to the total number of clock cycles.
An idle clock cycle is one where we saw the last signal of the previous column chunk but didn't see the first valid data beat of the next column chunk yet.
In general, decoder utilization rises with the number of worker threads as we produce more decoding jobs in parallel.
We can also see that having more streams submitting queries in parallel helps with decoder utilization since we can overlap decoding of one query with executing the CPU side of another operator pipeline more often.
Finally, we see that prefetching helps with decoder utilization since we can overlap Oasis decoding with the rest of the corresponding operator pipeline.
Decoder utilization caps out at $85\%$ since we also load other string columns through the bypass stream and we have other traffic like interrupts and configuration reads and writes going through the PCIe interconnect.
Overall, decoder utilization is governed by how many independent decoding jobs the database system can expose at any point in time.

\subsection{Discussion} 
Our Oasis column-chunk decoders exhibit excellent performance with throughput close to the interconnect speed with a single decoder for many columns.
Only for columns with low encoding ratio, the decompression is still the limiting factor even though we doubled Snappy decompression performance with our doubled datapath width and dual copy resolution per clock cycle.
In end-to-end query execution, we can overlap Parquet decoding fully with CPU query runtime for many queries, completely getting rid of the massive decoding overhead.
Our throughput benchmark shows that we can thus match the performance of DuckDB with its native Parquet extension and 16 worker threads with our Oasis SmartNIC on 8 worker threads.
The query throughput for Oasis is mainly limited by the PCIe interconnect.

The two residual overheads of the Oasis scan are string decoding and pushed-down filters that are still evaluated on the CPU.
These are the natural next targets to evolve the Oasis SmartNIC.
String decoding is a matter of extending the decoder pipeline with a variable-length dictionary that can output strings and a split output writer per decoder to write German string structs \cite{conf/cidr/NeumannF20} and the dictionary itself.
Implementing this would fully get rid of the Parquet decoding overhead leaving the PCIe interconnect as the bottleneck.
We can relax this bottleneck by pushing down filters which would avoid sending columns to main memory that are used purely for filtering and lower the pressure on the interconnect for all other columns by reducing them to the actually relevant data that pass the filter.
Together with interconnects such as PCIe 5 or CXL, this points towards scans over Parquet files on remote storage that cost no more than scans over in-memory data, and towards a division of labor in which the network datapath is responsible for turning storage formats into query-ready columns and the CPU or GPU is left with query processing alone \cite{journals/pvldb/KabicWDA25}.

%% file: sections/06_conclusion.tex
\section{Related Work}
Accelerating data ingestion in cloud-native database systems has been approached from several directions: redesigning the file formats themselves, offloading parsing and decoding to specialized hardware, processing data on SmartNICs \cite{conf/sigmod/Faghih0IB24} and DPUs \cite{journals/pvldb/GiouroukisNPZM25, conf/cidr/HuB0025}, and pushing computation into or closer to the storage layer.
OS2G \cite{conf/asplos/0008CLWFZZXLLZS25} moves the object-storage client onto a DPU and transfers data directly from the DPU into GPU memory, bypassing the host entirely.
Orthogonal to our work, there has been a lot of work on using FPGAs for individual database operators \cite{journals/vldb/FangMHLH20}, from sketches \cite{journals/pvldb/KieferPBM20} and integer vector compression \cite{conf/damon/MohsenMFB20} to LSM tree compaction \cite{conf/fast/ZhangWCXYHZH0CH20,conf/sigmod/HuangCWWHZLWCL19} and graph processing \cite{journals/pacmmod/YuTCCHW25}.

\labeltitle{Modern file formats}
A new generation of file formats replace general-purpose block compression with nested cascades of lightweight encodings~\cite{journals/pacmmod/KuschewskiSAL23, journals/pvldb/GienieczkoKNLG25} and organize data such that decoding maps efficiently onto the SIMD units of modern CPUs and GPUs~\cite{journals/pvldb/AfroozehB25}.
Complementary to faster decoding, aggressive file pruning reduces the amount of data that must be decoded in the first place based on statistics over the stored data~\cite{conf/sigmod/ZimmererDKWOK25}.
Careful engineering of the scan path narrows the gap without closing it: Rey et al.\ collect statistics for improved query optimization and data skipping during query evaluation~\cite{conf/btw/ReyFN23}.
A recent GPU-accelerated Presto reports reading Parquet an order of magnitude below the theoretical I/O rate of its infrastructure, attributes this to the continuous interpretation of Parquet's file-, row-group-, and page-level metadata during reads, and resorts to a simplified Arrow-like format to quantify the gap---while noting that the exabytes already stored in Parquet make migration impractical~\cite{journals/corr/abs-2606-24647}.
Oasis takes the opposite route: rather than changing the format to fit the compute, it changes the hardware to fit the format that data lakes and lakehouses actually use today.
Pruning remains fully complementary.

\labeltitle{Hardware-accelerated parsing and decoding} Offloading data ingestion to specialized hardware has already been deployed at cloud scale in the past: Redshift's AQUA was an FPGA-based acceleration layer placed between storage and compute~\cite{conf/sigmod/ArmenatzoglouBB22}.
In the research literature, offloading has been explored for a variety of formats.
On FPGAs, prior work demonstrates JSON-to-Arrow conversion~\cite{conf/fpt/PeltenburgHBMA21}, line-rate JSON parsing~\cite{conf/damon/DannW0FF22}, and, closest to our work, Parquet decoding~\cite{conf/icfpt/PeltenburgLH0AH20, conf/damon/KwonIRMF26}.
Peltenburg et al.\ specifically only cover a subset of the encoding and compression schemes found in practice.
More importantly, both remain isolated prototypes: they are evaluated as standalone decoders, leave open where the accelerator would sit in a real deployment, and are not integrated into a database system.
Oasis addresses all three gaps and most importantly integrates end-to-end into DuckDB so that decoding overlaps with query execution instead of merely being faster in isolation.

\labeltitle{Operator pushdown} An orthogonal approach to reducing scan pressure in the compute layer is to move computation closer to the data.
Pushing operators into the storage layer~\cite{conf/icde/YuYWGSAS20, journals/vldb/YangYSAS24} and inserting intermediate layers that partially evaluate queries~\cite{journals/debu/CaiGGNPPP18, conf/sigmod/ArmenatzoglouBB22} have both been explored successfully.
Commercial systems follow the same principle: Google BigQuery pushes filter and projection evaluation into its Capacitor storage layer~\cite{journals/pvldb/0001GLRSTVADMPS20}, and Oracle's Exadata offloads predicate filtering, column projection, and join Bloom filters to storage~\cite{exadata-smartscan}, in both cases significantly reducing the volume of data crossing the storage--compute boundary.

\section{Summary and Future Work}
We presented Oasis, a data-processing SmartNIC that offloads Parquet decoding into the network datapath between cloud object storage and the database system.
Oasis features a set of hardware column-chunk decoders handling decoding of real-world Parquet files, a software abstraction layer for buffer management and scheduling, and end-to-end integration into DuckDB via an extension.
Our evaluation shows that Oasis hides the cost of Parquet decoding in the datapath with minimal overhead, overlapping the decoding with query execution and almost doubling query throughput in the best case.
In future work, we will add string column decoding, an S3-compatible HTTP stack, and operator pushdown \cite{conf/damon/DannA26}.